\documentclass[aps,prl,twocolumn,email,superscriptaddress,groupedaddress,reprint]{revtex4}
\usepackage{natbib}
\usepackage{graphicx}
\usepackage{bbold}

\usepackage[final]{pdfpages}
\usepackage{pgffor}

\begin{document}

\title{Complete Polarization Control in Multimode Fibers with Polarization and Mode Coupling}

\author{Wen Xiong}
\affiliation{Department of Applied Physics,
	Yale University, New Haven, Connecticut 06520, USA}
\author{Chia Wei Hsu}
\affiliation{Department of Applied Physics,
	Yale University, New Haven, Connecticut 06520, USA}
\author{Yaron Bromberg}
\affiliation{Racah Institute of Physics, Hebrew University of Jerusalem, Jerusalem 91904, Israel}
\author{Jose Enrique Antonio-Lopez}
\affiliation{CREOL, The College of Optics and Photonics, University of Central Florida, Orlando, Florida 32816, USA}
\author{Rodrigo Amezcua Correa}
\affiliation{CREOL, The College of Optics and Photonics, University of Central Florida, Orlando, Florida 32816, USA}
\author{Hui Cao}
\email{hui.cao@yale.edu}
\affiliation{Department of Applied Physics,
	Yale University, New Haven, Connecticut 06520, USA}

\begin{abstract}
Multimode optical fibers have seen increasing applications in communication, imaging, high-power lasers and amplifiers. However, inherent imperfections and environmental perturbations cause random polarization and mode mixing, making the output polarization states very different from the input one. This poses a serious issue for employing polarization sensitive techniques to control light-matter interactions or nonlinear optical processes at the distal end of a fiber probe. Here we demonstrate a complete control of polarization states for all output channels by only manipulating the spatial wavefront of a laser beam into the fiber. Arbitrary polarization states for individual output channels are generated by wavefront shaping without constraint on input polarizations. The strong coupling between spatial and polarization degrees of freedom in a multimode fiber enables full polarization control with spatial degrees of freedom alone, transforming a multimode fiber to a highly-efficient reconfigurable matrix of waveplates.
\end{abstract}

\maketitle

\thispagestyle{plain} 

\section{Introduction}\label{sec:1}

The vectorial nature of electromagnetic waves plays an indispensable role in light-matter interaction, optical transmission and imaging. A control over the polarization state of light has been widely exploited in single molecule detection, nanoplasmonics, optical tweezers, nonlinear microscopy and optical coherence tomography. However, a well-prepared state of polarization can be easily scrambled by multiple scatterings of light in disordered media. The other side of the coin is that multiple scatterings couple spatial and polarization degrees of freedom, enabling polarization control of the scattered light via wavefront shaping of the incident beam. Arbitrary polarization states have been attained in a single or a few spatial channels \cite{guan2012polarization, park2012dynamic, tripathi2012vector, tripathi2014harnessing,de2015polarization}, transforming the random medium to a dynamic wave plate. Nevertheless, it is extremely difficult to control the polarization state of the {\it total} transmitted light, and the relatively low transmission through a scattering medium limits the efficiency. 

\begin{figure}[htbp]
	\centering
	\includegraphics[width=0.9\linewidth]{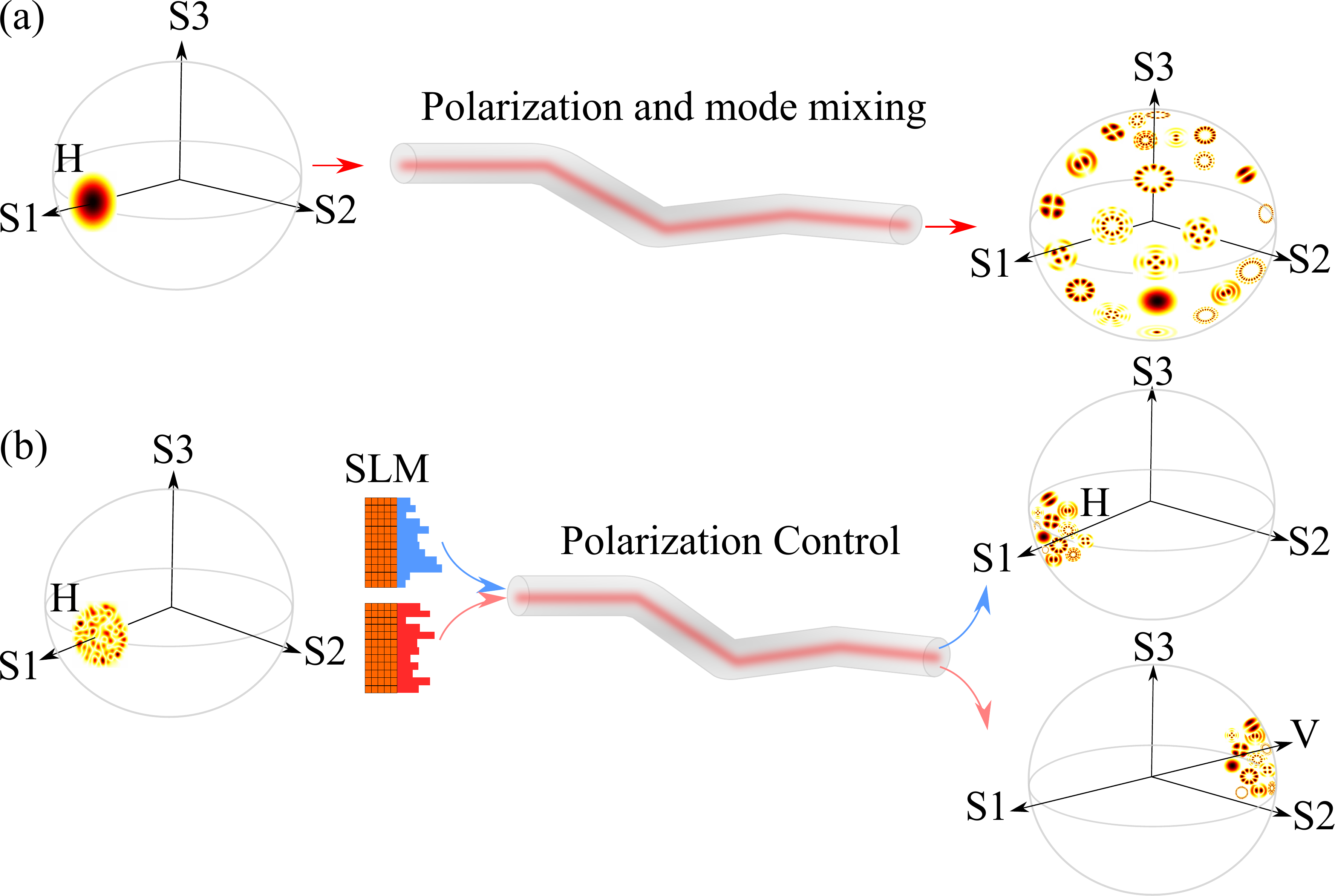}
	\caption{Fiber depolarization and polarization control by wavefront shaping. The three axes are the three Stokes parameters. 
		(a) Light is launched into the fundamental LP mode with the horizontal polarization, and subsequently coupled to other modes with different spatial profiles and polarization states while propagating in the fiber. The transmitted light is composed of all spatial modes in different polarization states, which are randomly spread over the Poincar{\'e} sphere. 		
		(b) Wavefront shaping of the horizontally polarized light by an SLM can overcome depolarization in the fiber, retaining the horizontal polarization for all output modes (top). A different input wavefront can convert all output modes to the vertical polarization (bottom).}
	\label{fig:concept}
\end{figure}

Polarization scrambling also occurs in optical fibers \cite{Kiesewetter2010polarisation}. For a single mode fiber (SMF), the output polarization state can be controlled by manipulating the input polarization. Due to refractive index fluctuations introduced by inherent imperfection and environmental perturbation such as eccentricity, bending and twisting, a multimode fiber (MMF) experiences not only polarization mixing but also mode mixing. When light is launched into  a single guided mode in the fiber, it will spread to other modes, each of which will experience distinct polarization scrambling. Thus the output polarization state of the modes varies from one mode to another [see Fig.~\ref{fig:concept}(a)], prohibiting a control of output polarization states in {\it all} modes by adjusting the input polarization of a single mode. One approach to complete polarization control is to measure the full transmission matrix of the MMF and invert it to find the vector fields to be injected into individual modes. This approach requires simultaneous control of both spatial and polarization degrees of freedom at the input, which is technically demanding.    

The coupling between spatial and polarization degrees of freedom in a MMF, as in a random scattering medium \cite{mosk2012controlling,rotter2017light}, opens the possibility of utilizing only spatial degrees of freedom of the input wave to control the polarization state of the output field. The key question is whether such a control would be complete, in the sense that not only arbitrary polarization state can be attained for total transmitted light regardless of the input polarization, but also each output mode may have a polarization state that differs from each other in a designed manner. If the complete polarization control can be achieved by only shaping the spatial wavefront of an input beam, it would be easier to realize experimentally, as most spatial light modulators (SLMs) operate for one polarization. A complete control of output polarization states is essential to applications of MMFs in endoscopy \cite{vcivzmar2011shaping,vcivzmar2012exploiting,choi2012scanner,caravaca2013real,gu2015design,ploschner2015seeing,sivankutty2016extended,porat2016widefield,caravaca2017single}, spectroscopy \cite{redding2012using,redding2014high,wan2015high}, microscopy \cite{brasselet2011polarization, stasio2016calibration}, nonlinear optics \cite{wright2015controllable,wright2016self}, quantum optics \cite{defienne2016two,israel2017quantum}, optical communication \cite{richardson2013space}, and fiber amplifiers \cite{doya2001optimized, michel2007selective, fridman2011principal,fridman2012modal,chen2017integrated,florentin2017shaping}. 

Here we demonstrate experimentally the ultimate polarization control of coherent light transmitted through a MMF with strong mode and polarization coupling. By modulating the spatial wavefront of a linearly polarized beam, depolarizations in the MMF is completely eliminated and the transmitted light retains the input polarization. Moreover, a complete conversion of the input polarization to its orthogonal counterpart or any polarization state is achieved. We further tailor the polarization states of individual output channels utilizing spatial degrees of freedom, without constraint on the input polarization state. Our theoretical analysis and numerical modeling illustrate that the full control of polarizations via spatial wavefront shaping is only possible when mode coupling occurs in the fiber. The random mode mixing, often unavoidable in a MMF, can be harnessed for functional advantages. A multimode fiber can function as a highly efficient reconfigurable matrix of waveplates, capable of converting arbitrary polarization states of the incident field to any desired polarization states via wavefront shaping techniques.

\section{Result}

\subsection{Mode coupling}\label{subsec:1}
To illustrate the critical role played by spatial mode coupling in polarization control, let us first consider a MMF with only polarization mixing but no mode mixing. Linearly polarized (LP) modes are the eigenmode of a perfect fiber under the weak guiding approximation \cite{okamoto2010fundamentals}. However, linear polarizations cannot be retained even in a perfect optical fiber \cite{Kiesewetter2010polarisation}. Further more, birefringence induced by fiber imperfections and perturbations changes the polarization state. Light injected into each LP mode effectively propagates through a distinctive set of wave plates with random orientations of their optical axes. Eventually different LP modes have different polarizations and the total output field becomes depolarized. In the absence of mode coupling, the MMF behaves like a bundle of uncoupled SMFs. It is impossible to control the output polarization of each mode without manipulating their individual input polarizations.  

With mode mixing in the fiber, spatial and polarization degrees of freedom become coupled. The output polarization state depends not only on the polarization but also on the spatial wavefront of the input field. For illustration, we consider a fiber with only two modes, each of which has two orthogonal polarization states. The incident light is monochromatic and linearly polarized in the horizontal direction. The field is $1$ for mode 1, and  $e^{i\theta}$ for mode 2. Without mode coupling, the relative phase $\theta$ between the two modes does not affect the output polarization state of either mode. However, with mode coupling, the output field of one mode also depends on the input field of the other. For example, the vertical polarization of mode 1 has contributions from (i) the field in mode 1 converted to the vertical polarization and (ii) the field in mode 2 that is coupled to the vertically polarized mode 1. The relative phase of these two contributions can be changed by varying $\theta$, resulting in a constructive or destructive interference which modifies the amplitude of the vertically polarized field in mode 1. This degree of freedom is effective only when there is mode mixing in the fiber. Compared to a fiber without mode coupling, more polarization states can be created at the output by adjusting the input wavefront. Mode mixing enables a polarization control utilizing spatial degrees of freedom, as illustrated schematically in Fig.~\ref{fig:concept}(b).  

\begin{figure}
	\centering
	\includegraphics[width=\linewidth]{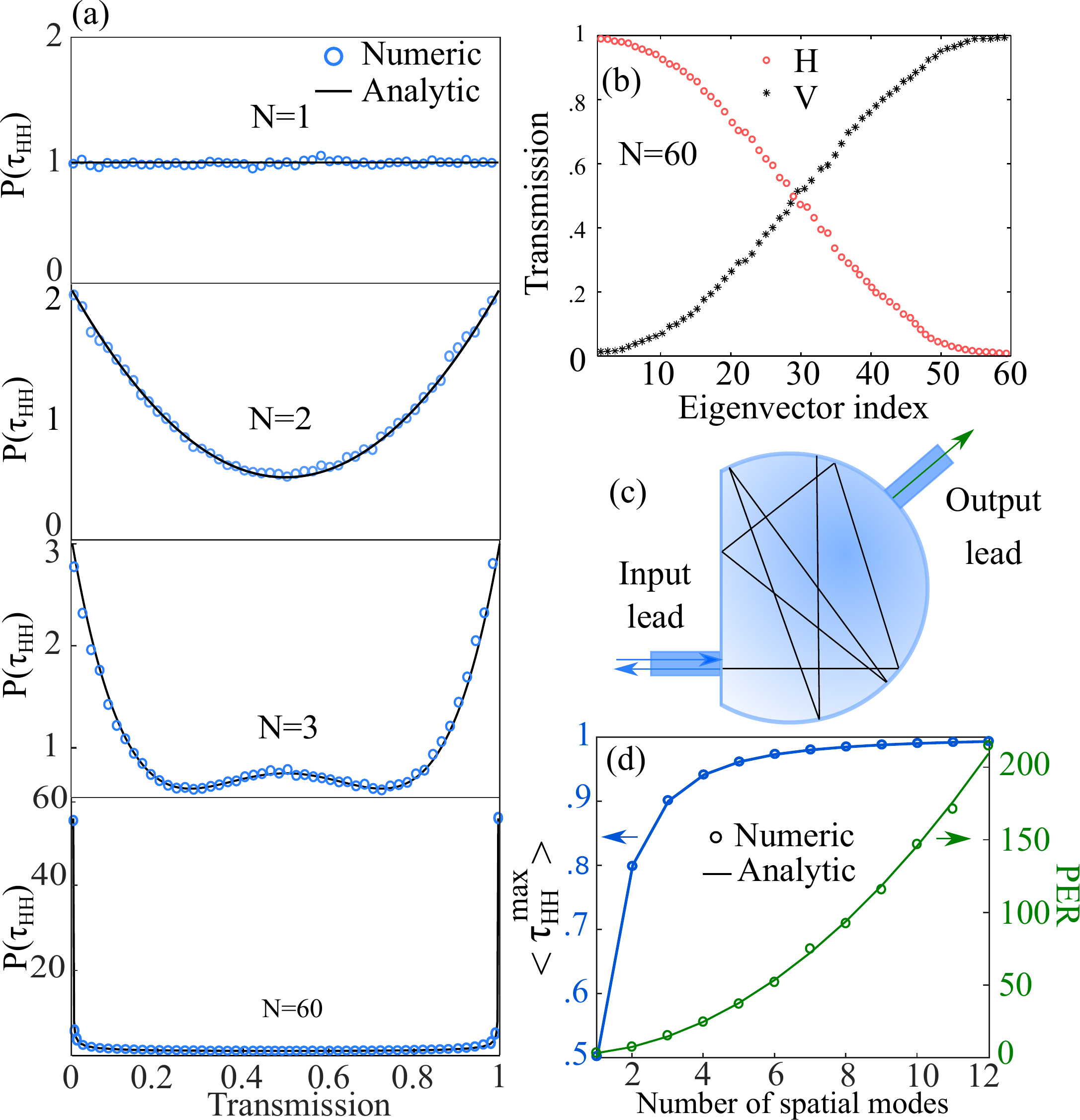}
	\caption{Polarization mixing in a multimode fiber and analogy to wave scattering in a chaotic cavity. 
		(a) Numerically calculated density of the eigenvalues of $t_{\rm HH}^\dagger t_{\rm HH}$ (blue circles) in a MMF with strong mode and polarization coupling but no loss. With increasing number of modes $N$ in the fiber, $P(\tau_{\rm HH})$ evolves to bimodal distribution, in agreement to the analytical expression for the reflection eigenvalue density in a chaotic cavity with two leads (black solid lines). 
		(b) Transmission of horizontal (H) and vertical (V) polarization components for individual eigenvectors of $t_{\rm HH}^{\dagger}t_{\rm HH}$, which are numbered by their eigenvalues from high to low. The decrease of H is accompanied by an increase of V, and their sum remains 1. 
		(c) Schematic diagram of a chaotic cavity with two leads. Wave enters the cavity through the input lead and undergoes multiple reflection from the cavity wall before exiting via the output lead (transmission) or the input lead (reflection). 
		(d) The maximum transmission of horizontal polarization, $\langle \tau^{\rm max}_{\rm HH} \rangle$, approaches 1 rapidly with increasing $N$. The polarization extinction ratio ($\rm PER$) scales as $N^2$. The symbols represent numerical data and the solid lines are analytic results.} 
	\label{fig:chaotic cavity}
\end{figure}

\subsection{Polarization Manipulation}
\subsubsection{Depolarization-free states}
To quantitatively evaluate the polarization control via spatial degrees of freedom only, we perform numerical simulation of a MMF with strong polarization and mode coupling. The fiber has $N$ modes, each of which has a two-fold degeneracy corresponding to two orthogonal polarizations. We use the concatenated fiber model \cite{ho2011statistics} to simulate random coupling among all modes of the MMF (see Supplementary Materials). Without loss of generality, we use the horizontal (H) and vertical (V) polarizations as the basis to describe the full transmission matrix of the MMF
\[
t=
\left[ {\begin{array}{cc}
	t_{\rm HH} & t_{\rm HV} \\
	t_{\rm VH} & t_{\rm VV} \\
	\end{array} } \right]
\] where $t_{\rm HH}$ ($t_{\rm VH}$) represents the horizontal (vertical) component of the output field when the input light is horizontally polarized. $t_{\rm HH}$ has the dimension of $N \times N$, where $N$ is the number of modes in the fiber for a single polarization.

The $N$ eigenvalues of $t_{\rm HH}^{\dagger}t_{\rm HH}$, denoted as $\tau_{\rm HH}$,  are in the range of 0 and 1. The largest eigenvalue determines the maximum energy that can be retained in the horizontal polarization  after propagating through the fiber. On the contrary, the smallest eigenvalue tells the maximum energy that can be converted to the vertical polarization. After simulating an ensemble of MMFs with random mode and polarization coupling, we obtain the eigenvalue density $P(\tau_{\rm HH})$ plotted in Fig.~\ref{fig:chaotic cavity}(a). If the fiber has only one mode ($N = 1$), $P(\tau_{\rm HH})$ has a uniform distribution between 0 and 1, as a result of the complete polarization mixing in the fiber. When there are two guided modes ($N = 2$), $P(\tau_{\rm HH})$ develops two peaks at $\tau_{\rm HH} = 0, 1$. With the increase of $N$, these two peaks grow rapidly and become dominant at $N \gg 1$. Thus the probability of finding extreme eigenvalues is very high. $\tau_{\rm HH} =  1$ means the transmitted light retain the input polarization (H), while $\tau_{\rm HH} =  0$ corresponds to 100$\%$ conversion to the orthogonal polarization (V). As $\tau_{\rm HH}$ decreases from 1 to 0, the percentage of transmission in the horizontal polarization drops, while that in the vertical polarization rises, as seen in Fig.~\ref{fig:chaotic cavity}(b).

The numerically calculated eigenvalue density $P(\tau_{\rm HH})$ agrees with the analytical prediction of the wave transmission in a lossless chaotic cavity [see the lines in Fig.~\ref{fig:chaotic cavity}(a)]. Such an agreement reveals the analogy between a MMF with random mode and polarization coupling and a chaotic cavity with two leads, as drawn schematically in Fig.~\ref{fig:chaotic cavity}(c). The transmission of the input polarization in the fiber is analogous to the reflection in the chaotic cavity in the sense that light exits the cavity via the same lead. Hence, the eigenvalue $\tau_{\rm HH}$ for the MMF corresponds to the reflection eigenvalue of the chaotic cavity. 

Using the analytical theory developed previously for the chaotic cavity \cite{baranger1994mesoscopic, jalabert1994universal}, we derive the probability density of the maximum eigenvalue of $t_{\rm HH}^{\dagger}t_{\rm HH}$ (see Supplementary Materials). The average value, $\big \langle\tau^{\rm max}_{\rm HH} \big \rangle = 1 - 1 / (N^2 + 1)$, approaches unity rapidly with the increase of $N$, in agreement with the numerical result shown in Fig.~\ref{fig:chaotic cavity}(d). The polarization extinction ratio (PER) of transmitted light is $\big \langle \tau_{\rm max} \big \rangle /(1 -  \big \langle \tau_{\rm max} \big \rangle) = N^2$. With a large number of modes in the fiber, we obtain ${\rm PER} \gg 1$. Depolarizations are avoided by coupling light into the eigenvector associated with the maximum eigenvalue of $t_{\rm HH}^{\dagger}t_{\rm HH}$. The eigenvector is a superposition of LP modes with the horizontal polarization, and can be generated by an SLM. The $N^2$ scaling originates from the repulsion between eigenvalues, which leads to the bimodal distribution of eigenvalues \cite{beenakker1997random, rotter2017light}. 

For comparison we consider the scaling of PER in a MMF without mode mixing. Due to distinctive polarization coupling for individual modes, the probability of retaining the input polarization for all output modes vanishes when the number of modes is large. The best solution to retain the input polarization is to only excite the mode with an output polarization closest to the input polarization. As detailed in Supplementary Materials, the PER of the transmitted light scales linearly with $N$, inferior to the $N^2$ scaling in the presence of strong mode coupling. This comparison shows that spatial mode mixing greatly enhances the ability of overcoming depolarization. 

\subsubsection{Polarization conversion}
The efficiency of converting the input polarization (H) to the orthogonal polarization (V) is given by the minimum eigenvalue of $t_{\rm HH}^{\dagger}t_{\rm HH}$. When loss in the MMF is negligible, the minimum eigenvalue of $t_{\rm HH}^{\dagger}t_{\rm HH}$ and the maximum eigenvalue of $t_{\rm VH}^{\dagger}t_{\rm VH}$ correspond to the same eigenvector, since $t_{\rm HH}^{\dagger}t_{\rm HH} + t_{\rm VH}^{\dagger}t_{\rm VH} = \mathbb{1}$. When the polarizations are completely mixed, the transmitted field has no memory of its initial polarization, so the transmission matrix $t_{\rm VH}$ has the same statistical property as $t_{\rm HH}$. The eigenvalue density $P(\tau_{\rm VH})$ is identical to $P(\tau_{\rm HH})$ and has a bimodal distribution. The ensemble-averaged value $\langle \tau^{\rm min}_{\rm HH} \rangle = 1 - \langle \tau^{max}_{\rm VH} \rangle = 1/(N^2+1)$, and ${\rm PER} = N^2$. In Supplementary Materials, we provide the analytical expression for its probability density $P(\tau^{\rm min}_{\rm HH})$. When $N \gg 1$, light is almost completely transformed into the orthogonal polarization by spatially coupling light into the eigenvector of $t_{\rm HH}^{\dagger}t_{\rm HH}$ with the minimum eigenvalue. 

If loss in the fiber is significant,  the bimodal distribution of eigenvalue density will be modified, and the peak at eigenvalue equal to 1 will be suppressed. However, the peak near 0 is robust against loss, ensuring a complete polarization control, though the total transmission is reduced. For example, if the input light is horizontally polarized, by coupling it to the eigenvector of $t_{\rm VH}^{\dagger}t_{\rm VH}$ with eigenvalue close to 0, the transmitted light has a vanishing vertical component. Thus the depolarization is avoided, but part of incident light is lost instead of being transmitted. Also the transmitted light can be converted to the vertical polarization by exciting the eigenvector of $t_{\rm HH}^{\dagger}t_{\rm HH}$ with the minimum eigenvalue. 

So far we have considered only horizontal and vertical polarizations for input and output fields, but the same concept applies to any polarization state. As long as the fiber completely scrambles the polarization of light, all polarization states are equivalent. For example, for a horizontally polarized input (H) and when examining its output in the right-hand circular polarization (R), the corresponding transmission matrix $t_{\rm RH}^{\dagger}t_{\rm RH}$ has the same eigenvalue density as $t_{\rm HH}^{\dagger}t_{\rm HH}$. With strong mode coupling and negligible loss in the fiber, $P(\tau_{\rm RH})$ is bimodal, the peak at $\tau_{\rm RH} = 1$ ($\tau_{\rm RH} = 0$) allows a full conversion of horizontal polarization to right (left) circular polarization.  

\subsubsection{Multi-channel polarization transformation}
\begin{figure}
	\centering
	\includegraphics[width=0.8\linewidth]{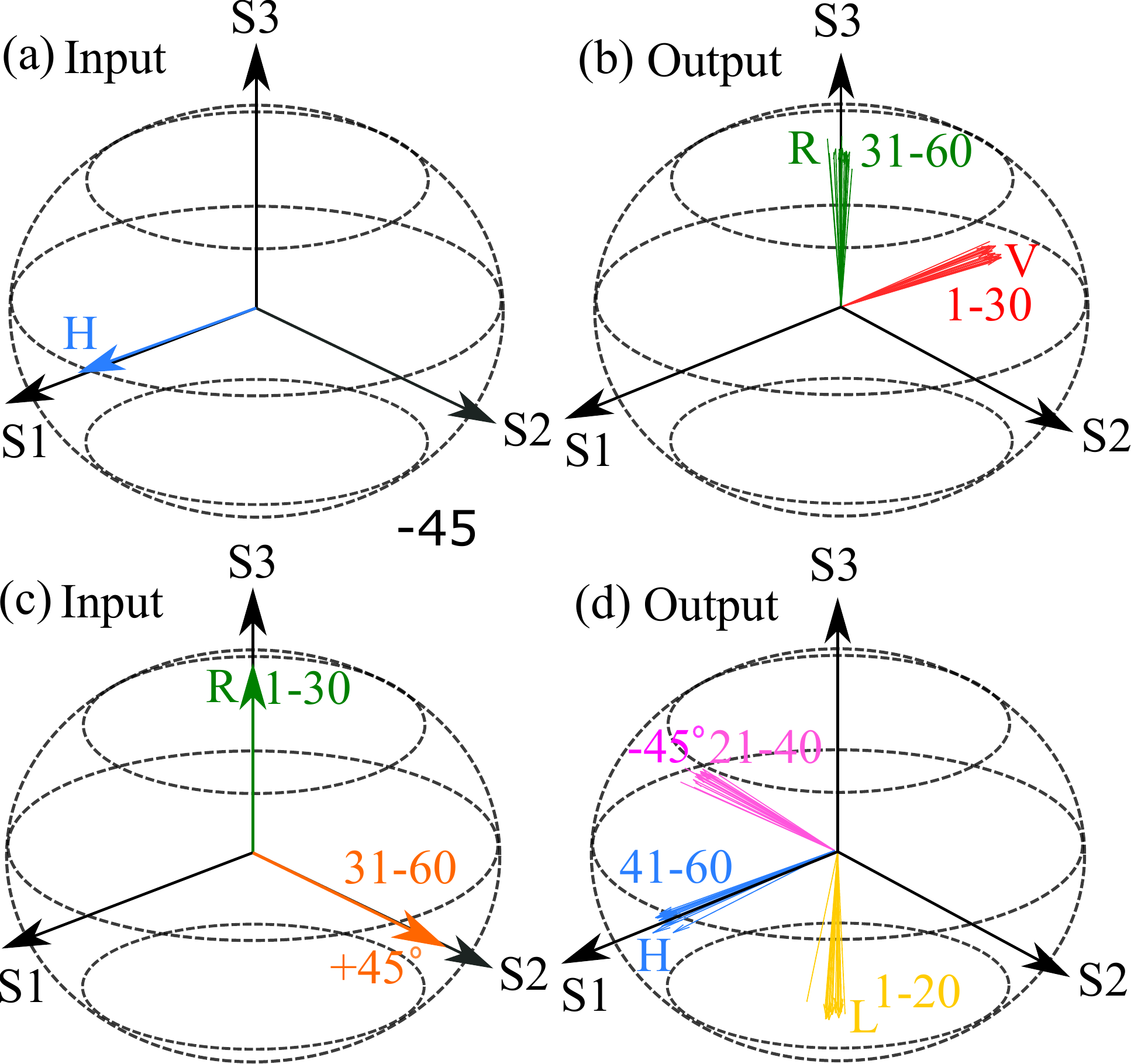}
	\caption{Poincar{\' e} sphere representation of multi-channel polarization transformation in the MMF. The direction of each arrow stands for the polarization of each mode and the length represents the intensity of the mode. (a, b) Transformation of the input horizontal polarization (H) to the output polarization state with the vertical polarization (V) for modes 1-30 and the right-hand circular polarization (R) for modes 31-60.
		(c, d) Transformation of the input polarization state with the right-hand circular polarization (R) for modes 1-30 and the linear +45 $^\circ$ polarization for modes 31-60 to the output polarization state with the left-hand circular polarization (L) for modes 1-20, the linear -45 $^\circ$ polarization for modes 21-40, and the horizontal polarization (H) for modes 41-60.
	}
	\label{fig:polarization}
\end{figure}

Let us take one step further: instead of controlling the polarization state of the total transmission, we can have different polarization states for different modes. As an example, we transform the horizontal polarization (H) of the input field [Fig.~\ref{fig:polarization} (a)] to a complex polarization state (A) at the output of a MMF with 60 modes. As shown in Fig.~\ref{fig:polarization} (b), the polarization state A has the vertical polarization (V) for modes 1-30 and right-hand circular polarization (R) for modes 31-60. The conversion is achieved by coupling the incident light to the eigenvector of $t_{\rm AH}^{\dagger}t_{\rm AH}$ with the eigenvalue close to 1, when the loss in the fiber is negligible. When the loss is significant, we resort to the output polarization state B that is orthogonal to A. In this case, B has the horizontal polarization (H) for modes 1-30 and the left-hand circular polarization (L) for modes 31-60. By exciting the eigenvector of $t_{\rm BH}^{\dagger}t_{\rm BH}$ with the eigenvalue close to 0, the output polarization state is orthogonal to B and thus identical to A.

Finally, we can also handle arbitrary input polarization states, i.e. individual spatial modes may have different polarizations. By adjusting the input spatial wavefront, arbitrary polarizations can be obtained at the output. Figure~\ref{fig:polarization} (c) and (d) present one example (see the caption). Therefore, a MMF with strong mode and polarization coupling is capable of transforming arbitrary input polarizations to arbitrary output polarizations with nearly 100$\%$ efficiency. Since only the spatial degrees of freedom are deployed at the input, the output intensity in each spatial mode, i.e., the distribution of output energy among spatial modes, cannot be controlled. To design not only polarizations but also intensities of all output modes, both spatial and polarization degrees of freedom at the input shall be utilized. 

\subsection{Experimental Demonstration}

\begin{figure}[t]
	\centering
	\includegraphics[width=\linewidth]{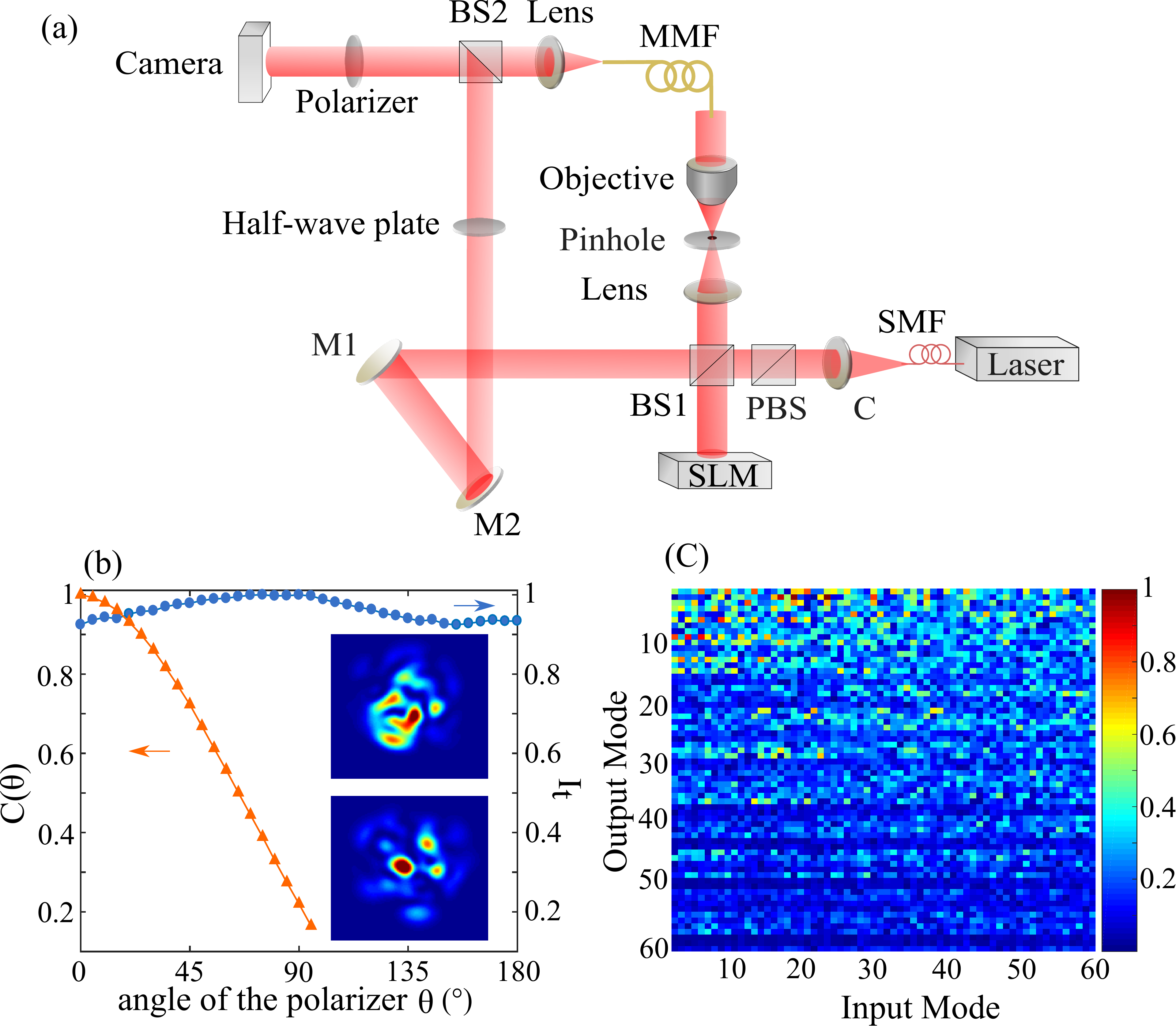}
	\caption{Experimental setup and fiber calibration. 
		(a) Schematic of the interferometric setup for measuring the transmission matrix of a multimode fiber (MMF). SMF: single mode fiber. C: lens. BS: beam splitter. PBS: polarizing beam splitter. M: mirror. 
		(b) Characterization of depolarization in the MMF. The total transmitted intensity $I_t$ (blue circles) and the correlation of output intensity patterns $C(\theta)$ (orange triangles) confirm complete depolarization. $\theta$ is the angle of the polarizer. 
		(c) Amplitude of measured transmission matrix $t_{\rm HH}$ in LP mode basis reveals strong mode mixing in the fiber.}
	\label{fig:setup}
\end{figure}

We experimentally demonstrate the complete polarization control of a MMF with strong polarization and mode coupling by wavefront shaping. We characterize the polarization-resolved transmission matrix with an interferometric setup shown in Fig.~\ref{fig:setup}(a). A detailed description is in \textbf{Materials and Methods}. To quantify the depolarization in the MMF, we measure the total transmitted intensity $I_t$ as a function of the angle of the polarizer $\theta$. As shown in Fig.~\ref{fig:setup}(b), $I_t$ only exhibits slight ($\sim 9 \%$) variations with $\theta$. Furthermore, the output intensity pattern changes with $\theta$ and thus individual output channels have distinct polarizations. We compute the correlation function $C(\theta) = \vec{I}(0) \cdot \vec{I}(\theta)$, where $\vec{I}(\theta)$ is a unit vector representing the normalized intensity profile at $\theta$. The decay of $C(\theta)$ in Fig.~\ref{fig:setup}(b) illustrates the decreasing correlation of the intensity pattern with $\theta$. The insets of Fig.~\ref{fig:setup}(b) are the two intensity patterns of orthogonal polarizations ($\theta = 0, 90^{\circ}$), which are almost uncorrelated, indicating nearly complete depolarization.

The amplitude of measured transmission matrix $t_{\rm HH}$ of the MMF in the LP mode basis is shown in Fig.~\ref{fig:setup}(c). No matter which mode light is injected into, the output field spreads over all modes, although higher order modes have lower amplitudes due to stronger dissipations. The measured $t_{\rm VH}$, given in Supplementary Materials, has a similar characteristic to that of $t_{\rm HH}$.  These results confirm strong spatial and polarization mixing in the MMF.

\begin{figure}
	\centering
	\includegraphics[width=\linewidth]{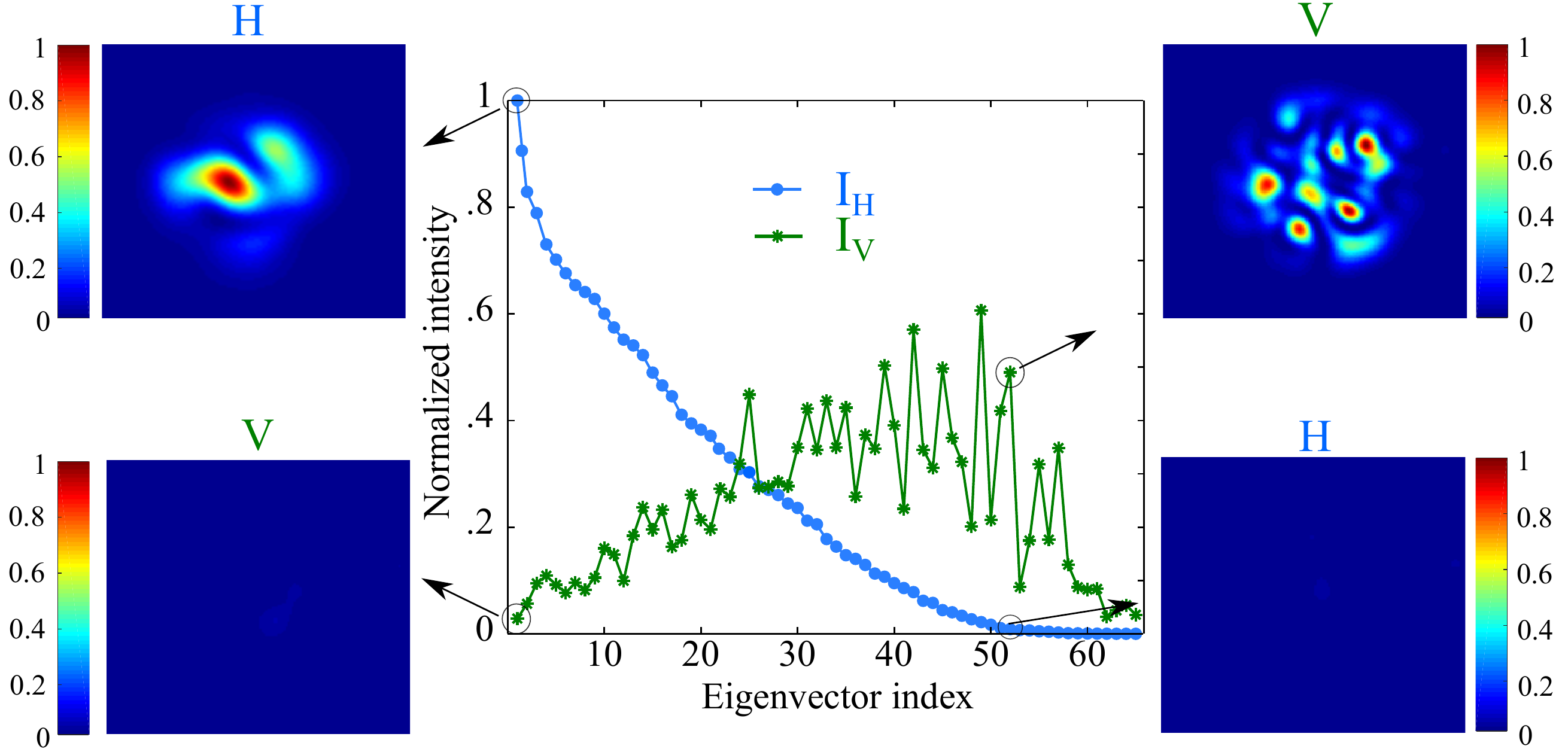}
	\caption{Experimental demonstrations of overcoming fiber depolarization and complete conversion to the orthogonal polarization. Central panel:  intensities of the horizontal (H) and vertical (V) polarization components of total transmitted light, $I_H$ and $I_V$, for individual eigenvectors of experimentally measured $t_{\rm HH}^{\dagger}t_{\rm HH}$. The eigenvectors are arranged by $I_H$ from high to low, and the largest value of $I_H$ is normalized to 1. The experimentally measured output intensity patterns of H and V for the 1st and the 52nd eigenvectors are shown on the left and right, respectively.}
	\label{fig:transmission}
\end{figure}

To control the output polarization, we compute the eigenvectors of the experimentally measured $t_{\rm HH}^{\dagger}t_{\rm HH}$. For each eigenvector, the intensities of horizontal and vertical polarization components in total transmitted light, $I_{\rm H}$ and $I_{\rm V}$, are plotted in Fig.~\ref{fig:transmission}. The decrease of $I_{\rm H}$ is accompanied by the increase of $I_{\rm V}$, eventually $I_{\rm V}$ cannot reach the maximum of $I_{\rm H}$ due to mode-dependent loss in the fiber. Employing the computer-generated phase hologram for a simultaneous phase and amplitude modulation \cite{arrizon2007pixelated}, we create the input wavefront for the first eigenvector with the SLM and launch it into the fiber. The output intensity patterns of horizontal and vertical polarizations are recorded (left panel in Fig.~\ref{fig:transmission}). Since higher order modes suffer more loss in the fiber , the transmitted light is mainly composed of lower order modes. The horizontal polarization component is much stronger than the vertical one, and the PER is about 24. Hence, most energy is retained in the input polarization (H), and depolarization is overcome.  

A complete conversion to the orthogonal polarization (V) is achieved with the eigenvector of  $t_{\rm HH}^{\dagger}t_{\rm HH}$ with a small eigenevalue. For example, we choose the 52nd eigenvector which has a low transmission of the horizontal polarization and launch its input field profile into the MMF. The measured output intensity patterns are shown in the right panel of Fig.~\ref{fig:transmission}, and the transmitted light is dominated by the vertical polarization component. The PER is 43, exceeding that of the first eigenvector. Since the 52nd eigenvector has more contributions from higher order modes, which experience a higher attenuation than lower order modes, its transmission is about half that of the first eigenvector.    

\begin{figure}
	\centering
	\includegraphics[width=\linewidth]{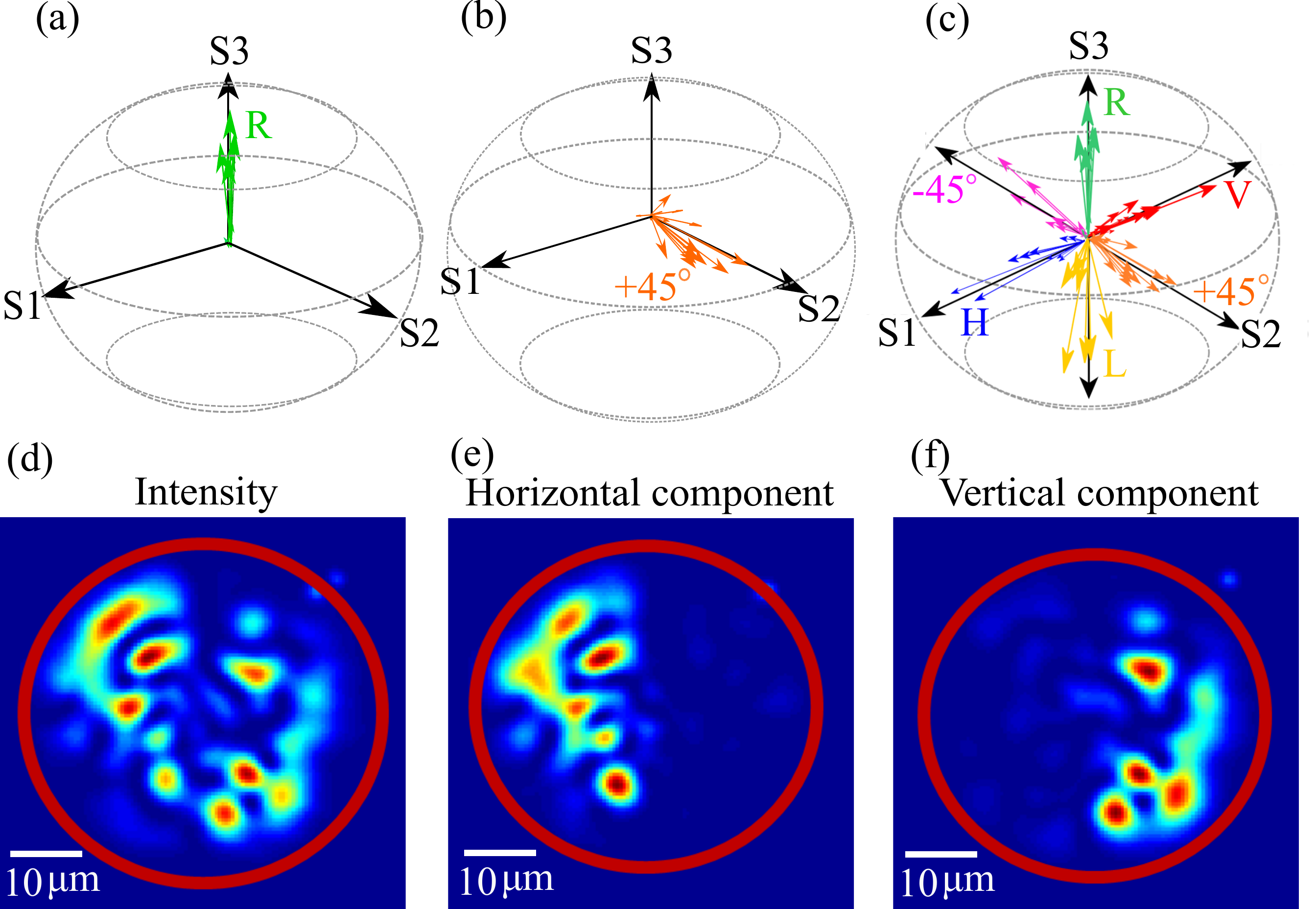}
	\caption{Experimental generation of arbitrary polarization states. 
		(a-c) Poincar{\' e} sphere representation of output polarization state of (a) the highest transmission eigenchannel of $t_{\rm RH}$,  all LP modes (each represented by an arrow) are right circularly polarized, (b) a low transmission eigenchannel of $t_{\rm FH}$, all modes are linearly polarized in the $45^{\circ}$ direction. (c) different polarization states generated with a fixed input polarization (H). 
		(d-f) Output intensity pattern (d), its horizontal (e) and vertical (f) polarization components reveal the transmitted field in the left half of the fiber facet is horizontally polarized, and the right half vertically polarized.}
	\label{fig:pattern}
\end{figure}

We can convert the horizontally polarized light to any polarization state at the fiber output. For example, to obtain the right circular polarization (R), we construct $t_{\rm RH} = \frac{1}{\sqrt{2}} (t_{\rm HH} - it_{\rm VH})$ from the measured $t_{\rm HH}$ and $t_{\rm VH}$, and couple the incident light to the eigenvector of $t^{\dagger}_{\rm RH} t_{\rm RH}$ with the largest eigenvalue. The output polarization states of individual LP modes are plotted on a Poincar\'e sphere in Fig.~\ref{fig:pattern}(a). Each arrow represents one mode, and its length stands for the intensity of that mode. All arrows point along the $S_3$ axis, indicating all modes are circularly polarized, despite varying intensities. In Fig.~\ref{fig:pattern}(b), we obtain the linear $+45^{\circ}$ polarization by exciting a low transmission eigenchannel of $t_{\rm -45H} = \frac{1}{\sqrt{2}} (t_{\rm HH} - t_{\rm VH})$ (the eigenvector of $t^{\dagger}_{\rm -45H} t_{\rm -45H}$ with a small eigenvalue). Figure~\ref{fig:pattern}(c) shows different polarization states that are generated experimentally with a fixed input polarization (H).

To demonstrate the ultimate polarization control, we make the polarization states different for individual output channels. In addition to the fiber mode basis, the spatial channels can be represented in real space (near-field zone of the fiber distal end) or momentum space (far-field zone). In the following example, we describe the fiber output channels in real space. The output polarization state C is designed to have the horizontal polarization for the spatial channels within the left half of the fiber cross-section, and the vertical polarization in the right half. The transmission matrix $t_{\rm CH}$ is constructed by concatenating one half of $t_{\rm HH}$ and the other half of $t_{\rm VH}$. The conversion of polarization from H to C is realized by exciting the highest transmission eigenchannel of $t_{\rm CH}$. Figure~\ref{fig:pattern}(d) is an image of the fiber output facet taken by the camera without a polarizer. After the linear polarizer is placed in front of the camera and oriented in the horizontal direction, the right half becomes dark while the left half remains bright in Fig.~\ref{fig:pattern}(e). Once the polarizer rotates to the vertical direction, the right half lights up while the left half turns dark  in Fig.~\ref{fig:pattern}(f). Hence, the transmitted field is horizontally polarized in the left half of the fiber facet, and vertically polarized in the right half. Additional example is given in Supplementary Materials, showing the output field is the left-hand circular polarization (L) in the top half of the fiber facet and right-hand circular polarization (R) in the bottom half. 

\section{Discussion}

We demonstrate that strong coupling between spatial and polarization degrees of freedom in a multimode fiber enables a complete control of output polarization states by manipulating only the spatial input wavefront. A general procedure of finding the spatial wavefront to create arbitrary polarization state is outlined and confirmed experimentally. It involves a measurement of the polarization-resolved transmission matrix and a selective excitation of the transmission eigenchannels corresponding to the extremal eigenvalues. With random mixing among all modes of different polarizations in the fiber, the probability of having extremal eigenvalues is enhanced by eigenvalue repulsion, analogous to a chaotic cavity. The global control of polarization states is not only useful for overcoming the depolarization in a multimode fiber, but also valuable for employing polarization-sensitive imaging techniques of fiber endoscopy and nonlinear microscopy. 

\section{Materials and Methods}
The core diameter of the fiber is 50 $\mu$m and the numeric aperture is approximately 0.22. It has a graded index profile designed to reduce mode dependent loss (see Supplement Materials). There are about 60 modes for a single polarization. We characterize the polarization-resolved transmission matrix with an interferometric setup shown in Fig.~\ref{fig:setup}(a). A horizontally polarized laser beam at wavelength $\lambda = 1550 $ nm is  split into a fiber arm and a reference arm. The SLM in the fiber arm prepares the spatial wavefront of light before it is launched into the multimode fiber. A half-waveplate in the reference arm rotates the polarization direction of the reference beam, which will combine with the transmitted light through the fiber at a beam-splitter. A linear polarizer in front of the camera removes the orthogonal polarization component of the output. From the interference fringes recorded by a camera, we extract the amplitude and phase of output field exiting the fiber with the same polarization as the reference \cite{xiong2016spatiotemporal}.

\section{Acknowledgements}
The authors thank H. Yilmaz, S. Gertler, S. Rotter and T. Kottos for stimulating discussions. This work is
supported by the U.S. National Science Foundation
under the Grants No. ECCS-1509361.

\section{Supplementary Materials}
Supplementary Material accompanies this paper. 
\clearpage


\section*{Supplementary material}

# Complete Polarization Control in Multimode Fibers with Polarization and Mode Coupling: Supplementary Information

Wen Xiong,[1] Chia Wei Hsu,[1] Yaron Bromberg,[2] Jose Enrique Antonio-Lopez,[3] Rodrigo Amezcua Correa,[3] and Hui Cao[1]

[1] *Department of Applied Physics, Yale University, New Haven, Connecticut 06520, USA*
[2] *Racah Institute of Physics, Hebrew University of Jerusalem, Jerusalem 91904, Israel*
[3] *CREOL, The College of Optics and Photonics, University of Central Florida, Orlando, Florida 32816, USA*

## CONCATENATED FIBER MODEL

We use the concatenated fiber model [1] to simulate a multimode fiber (MMF) with random mode and polarization coupling. The fiber is two-meter long and divided into 5 segments. The numerical aperture (NA) is 0.22. By varying the core diameter, we change the number of spatial modes. Each spatial mode has a two-fold degeneracy corresponding to horizontal (H) and vertical (V) linear polarizations. Light propagates without polarization or mode coupling in each segment. Between adjacent segments, all modes of different spatial profiles and polarizations are randomly coupled, which is modeled by a unitary random matrix. The total transmission matrix is the product of the transmission matrix for each segment.

## FIBER TRANSMISSION MATRIX

If loss in a multimode fiber (MMF) is negligible, the full transmission matrix for both polarizations

$$t = \begin{bmatrix} t_{\text{HH}} & t_{\text{HV}} \\ t_{\text{VH}} & t_{\text{VV}} \end{bmatrix}$$

is a random unitary matrix of dimension $2N \times 2N$, where $N$ is the number of spatial modes for a single polarization in the fiber. With strong polarization and mode coupling, such a matrix $t$ has no symmetry other than unitarity and is a member of the circular unitary ensemble (CUE) [2]. Since $t_{\text{HH}}$ and $t_{\text{VH}}$ are two statistically equivalent quarters of the full matrix $t$, they have identical statistical properties. Their eigenvalue density evolves to a bimodal distribution at large $N$.

## CHAOTIC CAVITY

The transmission matrix $t$ for a lossless MMF with strong polarization and mode coupling is mathematically analogous to the scattering matrix $s$ of a lossless chaotic cavity with two leads [see Fig.2 (c) of the main text]. Wave enters the chaotic cavity through one lead, then reflected multiple times from the cavity wall before escaping via the same lead or the other lead. Each lead is a waveguide with N statistically equivalent channels.

The four components of the scattering matrix

$$s = \begin{bmatrix} r_1 & t_2 \\ t_1 & r_2 \end{bmatrix}$$

correspond to transmissions and reflections at the two leads. When reciprocity is broken (e.g., via magnetic field), $t_1 \neq t_2$, and the $s$ matrix is a member of CUE. Both $r_{1,2}$ and $t_{1,2}$ are statistically equivalent $N \times N$ matrices. The density of transmission or reflection eigenvalues exhibits a bimodal distribution, $p(\tau) = 1/\pi \sqrt{\tau(1-\tau)}$, for large $N$ [3].

## MAXIMUM TRANSMISSION EIGENVALUE

The joint probability density for the $N$ eigenvalues of $t_{\text{HH}}^\dagger t_{\text{HH}}$ or $t_{\text{VH}}^\dagger t_{\text{VH}}$ for the MMF, $\{\tau_1, \ldots, \tau_N\}$, is identical to the joint probability of reflection or transmission eigenvalues of a chaotic cavity, which is [3, 4]

$$p(\tau_1, \ldots, \tau_N) = c_N \prod_{n<m} (\tau_n - \tau_m)^2, \quad (1)$$

with $\tau_n \in (0,1)$ for all $n$. Here $c_N$ is a normalization constant such that $\int (\prod_{n=1}^N d\tau_n) p(\tau_1, \ldots, \tau_N) = 1$ and $\prod_{n<m}$ is short for $\prod_{n=1}^N \prod_{m=n+1}^N$. The reduced probability when two eigenvalues are close by is a result of eigenvalue repulsion [2].

Let $\tau_{\max}$ be the largest among the $N$ eigenvalues. The probability density of $\tau_{\max}$ follows from (1) as

$$p(\tau_{\max}) = N \int_0^{\tau_{\max}} d\tau_1 \cdots \int_0^{\tau_{\max}} d\tau_{N-1} p(\tau_1, \ldots, \tau_{N-1}, \tau_{\max}). \quad (2)$$

The integrals in (2) gives

$$p(\tau_{\max}) = N^2 (\tau_{\max})^{N^2 - 1}, \quad \tau_{\max} \in (0,1). \quad (3)$$

It gives

$$\begin{aligned} \langle \tau_{\max} \rangle &= 1 - \frac{1}{N^2 + 1}, \\ \text{var}(\tau_{\max}) &= \frac{N^2}{(N^2 + 1)^2 (N^2 + 2)}. \end{aligned} \quad (4)$$

We see that both $1 - \langle \tau_{\max} \rangle$ and the standard deviation of $\tau_{\max}$ scale as $1/N^2$ for large $N$. This is

because the eigenvalues near 1 are pushed further toward 1 by the repulsion from the smaller eigenvalues and there are no eigenvalues larger than 1 to counter balance this push. We define the polarization extinction ratio (PER) as the maximal ratio of the transmissions in the two polarizations, $\langle\tau_{\max}\rangle/(1 - \langle\tau_{\max}\rangle) = N^2$. Similarly, the probability density for the smallest eigenvalue is $p(\tau_{\min}) = N^2(1 - \tau_{\min})^{N^2-1}$.

## POLARIZATION CONTROL WITHOUT MODE MIXING

With negligible mixing between different spatial modes in the fiber, each pair of LP modes with same spatial profile and orthogonal polarization is coupled due to birefringence. Light injected to a spatial mode experiences successive polarization rotations while propagating in the fiber. Each mode is depolarized differently with a random output polarization state, whose overlap with the desired output polarization is a random number between 0 and 1. When light is launched into all spatial modes of the fiber, the probability for all of them to have a substantial overlap with the desired polarization state at the output decreases exponentially with N. Therefore, simultaneous control of the polarization states of all modes cannot be achieved by using spatial degrees of freedom alone.

To have the transmitted light in a desired polarization state, one should inject light only to the mode whose output polarization is the closest to the desired one. This reasoning can be shown mathematically. Since $t_{\text{HH}}$ is now a diagonal matrix, the eigenvalues of $t_{\text{HH}}^\dagger t_{\text{HH}}$ are simply $N$ independent random numbers with the $n$-th eigenvalue being the overlap between the output of the $n$-th mode and the desired polarization state. The eigenvector with the largest eigenvalue corresponds to sending light only into the mode whose output polarization has the largest overlap with the desired polarization. The joint probability density of the $N$ eigenvalues is simply $p(\tau_1, \ldots, \tau_N) = \prod_{n=1}^{N} p(\tau_n) = 1$, and the probability density of the maximal eigenvalue [as given by Eq. (S2)] is

$$p\prime(\tau_{\max}) = N \int_0^{\tau_{\max}} p(\tau_1) d\tau_1 \cdots \int_0^{\tau_{\max}} p(\tau_{N-1}) d\tau_{N-1}. \quad (5)$$

It follows that

$$p\prime(\tau_{\max}) = N(\tau_{\max})^{N-1}, \quad \tau_{\max} \in (0, 1). \quad (6)$$

From it we have that $\langle\tau_{\max}\rangle = 1 - 1/(N+1)$ and PER $= \langle\tau_{\max}\rangle/(1 - \langle\tau_{\max}\rangle) = N$.

## FIBER REFRACTIVE INDEX PROFILE

The refractive index profile of the MMF tested in the experiment is designed to reduce mode-dependent loss. The core diameter is 50 $\mu$m and the measured refractive index profile is plotted in Fig. 1. A sharp drop of the refractive index at the interface between the core and the cladding enhances optical confinement and reduces the probability of light escaping from the core.

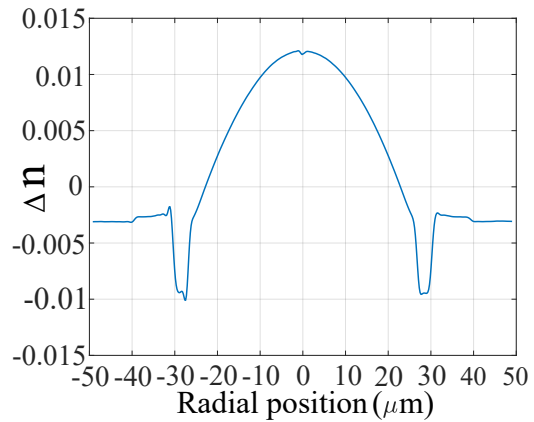

FIG. 1: Measured refractive index profile of the multimode fiber. The difference between the refractive index in the core and that in the cladding, $\Delta n$, has a parabolic profile within the core (from -25 $\mu$m to 25 $\mu$m), and a sharp drop at the interface between the core and the cladding to reduce light leakage.

## MEASURED TRANSMISSION MATRICES

Figure 2 presents experimentally measured field transmission matrices, $t_{\text{HH}}$ and $t_{\text{VH}}$, of the MMF for two output polarizations H and V, with the input polarization set to H. The amplitudes of matrix elements of $t_{\text{HH}}$ and $t_{\text{VH}}$ reveal strong mode and polarization mixing. The phases of all matrix elements are random, between 0 and $2\pi$, indicating full coupling among all spatial modes of different polarizations. The eigenvalues of $t_{\text{H}}^\dagger t_{\text{H}}$, in which $t_{\text{H}} = \begin{pmatrix} t_{\text{HH}} \\ t_{\text{VH}} \end{pmatrix}$, reveals the mode-dependent loss in the fiber. The sharp drop of the transmission curve after mode 50 indicates the cut-off of guided modes in the fiber, as shown in Fig. 3. The coupling is also strong when we use left-hand circular and right-hand circular polarizations as the basis.

## POLARIZATION CONTROL

Figure 4 shows an numerically generated polarization state using the experimentally measured transmission matrices $t_{HH}$ and $t_{VH}$. All spatial channels in the top half of the fiber facet are left-hand circularly polarized (L) and the bottom half right-hand circularly polarized (R).



\begin{thebibliography}{10}
	\newcommand{\enquote}[1]{``#1''}
	
	\bibitem{guan2012polarization}
	Y.~Guan, O.~Katz, E.~Small, J.~Zhou, and Y.~Silberberg, \enquote{Polarization
		control of multiply scattered light through random media by wavefront
		shaping,} Opt. Lett. \textbf{37}, 4663--4665 (2012).
	
	\bibitem{park2012dynamic}
	J.-H. Park, C.~Park, H.~Yu, Y.-H. Cho, and Y.~Park, \enquote{Dynamic active
		wave plate using random nanoparticles,} Opt. Express \textbf{20},
	17010--17016 (2012).
	
	\bibitem{tripathi2012vector}
	S.~Tripathi, R.~Paxman, T.~Bifano, and K.~C. Toussaint, \enquote{Vector
		transmission matrix for the polarization behavior of light propagation in
		highly scattering media,} Opt. Express \textbf{20}, 16067--16076 (2012).
	
	\bibitem{tripathi2014harnessing}
	S.~Tripathi and K.~C. Toussaint, \enquote{Harnessing randomness to control the
		polarization of light transmitted through highly scattering media,} Opt.
	Express \textbf{22}, 4412--4422 (2014).
	
	\bibitem{de2015polarization}
	H.~B. de~Aguiar, S.~Gigan, and S.~Brasselet, \enquote{Polarization-resolved
		microscopy through scattering media via wavefront shaping,} arXiv
	preprint:1511.02347  (2015).
	
	\bibitem{Kiesewetter2010polarisation}
	D.~V.~Kiesewetter, \enquote{Polarisation characteristics of light from multimode optical fibres,} 
	Quantum Electron. \textbf{40}, 519--524 (2010).
	
	\bibitem{mosk2012controlling}
	A.~P. Mosk, A.~Lagendijk, G.~Lerosey, and M.~Fink, \enquote{Controlling waves
		in space and time for imaging and focusing in complex media,} Nat. Photon.
	\textbf{6}, 283--292 (2012).
	
	\bibitem{rotter2017light}
	S.~Rotter and S.~Gigan, \enquote{Light fields in complex media: Mesoscopic
		scattering meets wave control,} Rev. Mod. Phys. \textbf{89}, 015005 (2017).
	
	\bibitem{vcivzmar2011shaping}
	T.~{\v{C}}i{\v{z}}m{\'a}r and K.~Dholakia, \enquote{Shaping the light
		transmission through a multimode optical fibre: complex transformation
		analysis and applications in biophotonics,} Opt. Express \textbf{19},
	18871--18884 (2011).
	
	\bibitem{vcivzmar2012exploiting}
	T.~{\v{C}}i{\v{z}}m{\'a}r and K.~Dholakia, \enquote{Exploiting multimode
		waveguides for pure fibre-based imaging,} Nat. Commun. \textbf{3}, 1027
	(2012).
	
	\bibitem{choi2012scanner}
	Y.~Choi, C.~Yoon, M.~Kim, T.~D. Yang, C.~Fang-Yen, R.~R. Dasari, K.~J. Lee, and
	W.~Choi, \enquote{Scanner-free and wide-field endoscopic imaging by using a
		single multimode optical fiber,} Phys. Rev. Lett. \textbf{109}, 203901
	(2012).
	
	\bibitem{caravaca2013real}
	A.~M. Caravaca-Aguirre, E.~Niv, D.~B. Conkey, and R.~Piestun,
	\enquote{Real-time resilient focusing through a bending multimode fiber,}
	Opt. Express \textbf{21}, 12881--12887 (2013).
	
	\bibitem{gu2015design}
	R.~Y. Gu, R.~N. Mahalati, and J.~M. Kahn, \enquote{Design of flexible
		multi-mode fiber endoscope,} Opt. Express \textbf{23}, 26905--26918 (2015).
	
	\bibitem{ploschner2015seeing}
	M.~Pl{\"o}schner, T.~Tyc, and T.~{\v{C}}i{\v{z}}m{\'a}r, \enquote{Seeing
		through chaos in multimode fibres,} Nat. Photon. \textbf{9}, 529--535 (2015).
	
	\bibitem{sivankutty2016extended}
	S.~Sivankutty, V.~Tsvirkun, G.~Bouwmans, D.~Kogan, D.~Oron, E.~R. Andresen, and
	H.~Rigneault, \enquote{Extended field-of-view in a lensless endoscope using
		an aperiodic multicore fiber,} Opt. Lett. \textbf{41}, 3531--3534 (2016).
	
	\bibitem{porat2016widefield}
	A.~Porat, E.~R. Andresen, H.~Rigneault, D.~Oron, S.~Gigan, and O.~Katz,
	\enquote{Widefield lensless imaging through a fiber bundle via speckle
		correlations,} Opt. Express \textbf{24}, 16835--16855 (2016).
	
	\bibitem{caravaca2017single}
	A.~M. Caravaca-Aguirre and R.~Piestun, \enquote{Single multimode fiber
		endoscope,} Opt. Express \textbf{25}, 1656--1665 (2017).
	
	\bibitem{redding2012using}
	B.~Redding and H.~Cao, \enquote{Using a multimode fiber as a high-resolution,
		low-loss spectrometer,} Opt. Lett. \textbf{37}, 3384--3386 (2012).
	
	\bibitem{redding2014high}
	B.~Redding, M.~Alam, M.~Seifert, and H.~Cao, \enquote{High-resolution and
		broadband all-fiber spectrometers,} Optica \textbf{1}, 175--180 (2014).
	
	\bibitem{wan2015high}
	N.~H. Wan, F.~Meng, T.~Schr{\"o}der, R.-J. Shiue, E.~H. Chen, and D.~Englund,
	\enquote{High-resolution optical spectroscopy using multimode interference in
		a compact tapered fibre,} Nat. Commun. \textbf{6} (2015).
	
	\bibitem{brasselet2011polarization}
	S.~Brasselet, \enquote{Polarization-resolved nonlinear microscopy: application
		to structural molecular and biological imaging,} Adv. Opt. Photonics
	\textbf{3}, 205 (2011).
	
	\bibitem{stasio2016calibration}
	N.~Stasio, C.~Moser, and D.~Psaltis, \enquote{Calibration-free imaging through
		a multicore fiber using speckle scanning microscopy,} Opt. Lett. \textbf{41},
	3078--3081 (2016).
	
	\bibitem{wright2015controllable}
	L.~G. Wright, D.~N. Christodoulides, and F.~W. Wise, \enquote{Controllable
		spatiotemporal nonlinear effects in multimode fibres,} Nat. Photon.
	\textbf{9}, 306--310 (2015).
	
	\bibitem{wright2016self}
	L.~G. Wright, Z.~Liu, D.~A. Nolan, M.-J. Li, D.~N. Christodoulides, and F.~W.
	Wise, \enquote{Self-organized instability in graded-index multimode fibre,}
	Nat. Photon. \textbf{10} (2016).
	
	\bibitem{defienne2016two}
	H.~Defienne, M.~Barbieri, I.~A. Walmsley, B.~J. Smith, and S.~Gigan,
	\enquote{Two-photon quantum walk in a multimode fiber,} Sci. Adv. \textbf{2},
	e1501054 (2016).
	
	\bibitem{israel2017quantum}
	Y.~Israel, R.~Tenne, D.~Oron, and Y.~Silberberg, \enquote{Quantum correlation
		enhanced super-resolution localization microscopy enabled by a fibre bundle
		camera,} Nat. Commun. \textbf{8} (2017).
	
	\bibitem{richardson2013space}
	D.~Richardson, J.~Fini, and L.~Nelson, \enquote{Space-division multiplexing in
		optical fibres,} Nat. Photon. \textbf{7}, 354--362 (2013).
	
	\bibitem{doya2001optimized}
	V.~Doya, O.~Legrand, and F.~Mortessagne, \enquote{Optimized absorption in a
		chaotic double-clad fiber amplifier,} Opt. Lett. \textbf{26}, 872--874
	(2001).
	
	\bibitem{michel2007selective}
	C.~Michel, V.~Doya, O.~Legrand, and F.~Mortessagne, \enquote{Selective
		amplification of scars in a chaotic optical fiber,} Phys. Rev. Lett.
	\textbf{99}, 224101 (2007).
	
	\bibitem{fridman2011principal}
	M.~Fridman, M.~Nixon, M.~Dubinskii, A.~A. Friesem, and N.~Davidson,
	\enquote{Principal modes in fiber amplifiers,} Opt. Lett. \textbf{36},
	388--390 (2011).
	
	\bibitem{fridman2012modal}
	M.~Fridman, H.~Suchowski, M.~Nixon, A.~A. Friesem, and N.~Davidson,
	\enquote{Modal dynamics in multimode fibers,} J. Opt. Soc. Am. A \textbf{29},
	541--544 (2012).
	
	\bibitem{chen2017integrated}
	H.~Chen, C.~Jin, B.~Huang, N.~Fontaine, R.~Ryf, K.~Shang, N.~Gr{\'e}goire,
	S.~Morency, R.-J. Essiambre, G.~Li \emph{et~al.}, \enquote{Integrated
		cladding-pumped multi-core, few-mode erbium-doped fibre amplifier for
		space-division multiplexed communications,} Nat. Photon. \textbf{10},
	529–533 (2016).
	
	\bibitem{florentin2017shaping}
	R.~Florentin, V.~Kermene, J.~Benoist, A.~Desfarges-Berthelemot, D.~Pagnoux,
	A.~Barth{\'e}l{\'e}my, and J.-P. Huignard, \enquote{Shaping the light
		amplified in a multimode fiber,} Light Sci. Appl. \textbf{6} (2017).
	
	\bibitem{okamoto2010fundamentals}
	K.~Okamoto, \emph{Fundamentals of optical waveguides} (Academic press, 2010).
	
	\bibitem{ho2011statistics}
	K.~P. Ho and J.~M. Kahn, \enquote{Statistics of group delays in multimode fiber
		with strong mode coupling,} J. Lightwave Technol. \textbf{29}, 3119--3128
	(2011).
	
	\bibitem{baranger1994mesoscopic}
	H.~U. Baranger and P.~A. Mello, \enquote{Mesoscopic transport through chaotic
		cavities: A random s-matrix theory approach,} Phys. Rev. Lett. \textbf{73},
	142 (1994).
	
	\bibitem{jalabert1994universal}
	R.~Jalabert, J.-L. Pichard, and C.~Beenakker, \enquote{Universal quantum
		signatures of chaos in ballistic transport,} EPL (Europhysics Letters)
	\textbf{27}, 255 (1994).
	
	\bibitem{beenakker1997random}
	C.~W. Beenakker, \enquote{Random-matrix theory of quantum transport,} Rev. Mod.
	Phys. \textbf{69}, 731 (1997).
	
	\bibitem{xiong2016spatiotemporal}
	W.~Xiong, P.~Ambichl, Y.~Bromberg, B.~Redding, S.~Rotter, and H.~Cao,
	\enquote{Spatiotemporal control of light transmission through a multimode
		fiber with strong mode coupling,} Phys. Rev. Lett. \textbf{117}, 053901
	(2016).
	
	\bibitem{arrizon2007pixelated}
	V.~Arriz{\'o}n, U.~Ruiz, R.~Carrada, and L.~A. Gonz{\'a}lez, \enquote{Pixelated
		phase computer holograms for the accurate encoding of scalar complex fields,}
	J. Opt. Soc. Am. A \textbf{24}, 3500--3507 (2007).
	
\end{thebibliography}
\end{document}